\documentclass[pra,aps,eqsecnum,showpacs, twocolumn]{revtex4}
\usepackage{mathtext}
\usepackage[T2A]{fontenc}
\usepackage[cp1251]{inputenc}
\usepackage{epsf}
\usepackage{psfrag}
\usepackage{graphicx}
\usepackage{amssymb,amsmath}

\begin{document}

\title{Many-body cooperative effects in an ensemble of point-like impurity centers near a charged conductive surface}
\author{A. S. Kuraptsev and I. M. Sokolov
\\
{\small Peter the Great St. Petersburg Polytechnic University, 195251, St. Petersburg, Russia}\\
}
\date{\today}

\sloppy



\begin{abstract}
On the basis of a quantum microscopic approach we study the cooperative effects induced by the dipole-dipole interaction in an ensemble of point-like impurity centers located near a charged perfectly conducting surface. We analyze the simultaneous influence of the modified spatial structure of field modes near the conductive surface and the electric field on the transition spectrum of an excited atom inside an ensemble and on the radiation trapping. We show that the electric field modifies the cooperative Lamb shift, as well as the character of sub- and superradiant decay. We also demonstrate that these modifications differ from those taking place in the case of atomic ensembles in free space, without conducting surface.
\end{abstract}


\maketitle
\section{Introduction}
Since the seminal work of Purcell \cite{1}, the interaction of light
with atoms localized inside a cavity or waveguide, as well as
near its surface, has attracted considerable attention. Now it is
well understood that a cavity modifies the spatial structure of the
modes of the electromagnetic field. This leads to the modification
of the radiative properties of atoms, and in particular to
the enhancement and inhibition of the spontaneous decay rate
\cite{2} -- \cite{5}. This proposes an exciting tool for the preparation of media with given optical properties. For this reason, the study of atomic systems in the presence of a cavity or waveguide is one of modern trends in atomic optics and quantum optics. Light interaction with atoms coupled to nanophotonic
structures, such as nanofibers \cite{6} -- \cite{8}, photonic crystal
cavities \cite{9} and waveguides \cite{10} -- \cite{11}, may have future applications
in quantum metrology, scalable quantum networks and
quantum information science \cite{12} -- \cite{14}.

Modification in the structure of field modes changes not only single-particle characteristics but also the nature of photon exchange between different atoms. In its turn this leads to an alteration of the dipole–dipole interatomic
interaction \cite{15} -- \cite{16}, as well as associated cooperative
effects \cite{17} -- \cite{18}. Cooperative properties of cold atomic gases coupled with dielectric nanoscale structures, in particular, nanofiber \cite{19} -- \cite{24} and photonic crystals \cite{25} -- \cite{27} are discussed nowadays.

In fact, not only cavity or waveguide can modify the spatial structure of the modes of the electromagnetic
field. Single metallic surface also has this property. For this
reason, the characteristics of the ensemble of atoms or quantum dots located
near the conductive surface differ from ones in the case of
the same ensemble in free space \cite{31}. If the metallic surface is charged, an electrostatic field causes Stark shifts of the atomic energy levels, which leads to additional modification of the interatomic dipole–dipole interaction \cite{KS_Laser_Phys_2018} -- \cite{KS_JETP_2018}. However, many-body cooperative effects, including multiple and recurrent light scattering, Anderson localization of light, induced by the dipole–dipole interaction in an ensemble of point-like impurity centers near a charged conductive surface, have not been studied in detail yet.

The goal of this paper is to describe theoretically polyatomic cooperative effects in a dense ensemble of point-like impurity
centers embedded in a solid dielectric and placed near a perfectly conductive charged plate. We simultaneously analyze two factors affecting the character of cooperative effects in the system with strong interatomic correlations: the peculiarities of the spatial structure of field modes near the conductive surface as well as Stark splitting of energy levels induced by an electrostatic field.  We show that the influence of the electric field on the collective effects in a dense polyatomic ensemble located near the conductive surface significantly differs from that in the case of an ensemble in free space, without a surface.

\section{Basic assumptions and approach}
Let us consider an ensemble which consists of $N$ motionless impurity atoms imbedded into transparent dielectric and
placed near a charged perfectly conducting plate. The longitudinal sizes of the plate are assumed to be significantly larger
than resonant transition wavelength $\lambda_{0}$ and the sizes of the atomic sample. We will suppose also that the temperature of the system is low enough to neglect the electron-phonon interaction. In this case the influence of the dielectric matrix on impurity atoms is restricted by
random shifts of their energy levels caused by inhomogeneous internal fields in dielectric. These approximations allow us to consider dynamics of the model system consisting of the set of motionless pointlike scatterers and the electromagnetic field.

At the present time, there are several approaches to the description of collective effects in the system under consideration  \cite{foldy45, lax51, 30a,31a,32a,33a,34a,35a,36a,SKH_JETP_2011,38a,39a}. In this work we use the consistent quantum-posed theoretical approach. In the framework of this approach, the considered quantum system is described by the wave function, which can be found by the method proposed first in \cite{HM, Heitler} and developed afterward in \cite{SKH_JETP_2011} for a description of the collective effects in dense and cold nondegenerate atomic gases. This method
was successfully used for the analysis of the optical properties of dense atomic ensembles as well as for studying light scattering from such ensembles \cite{KS_J_Mod_Opt_2013} -- \cite{KS_PRA_2017}.

Further this method was generalized on the case of atomic systems located in a Fabry-Perot cavity \cite{KS_JETP_2016, KS_PRA_2016}. In  the papers \cite{KS_Laser_Phys_2018} and \cite{KS_JETP_2018} it was used to analyze the dipole–dipole interaction between two motionless point atoms near a single perfectly conducting mirror.

The quantum microscopic approach was described at great length in several our papers \cite{SKH_JETP_2011, KS_PRA_2016, KS_JETP_2018} and we will not reproduce the general theory in detail  here. In the following paragraphs, we just provide a brief overview of it. The reader is referred to the mentioned papers for the theoretical developments and justifications.

The method employed is based on the solution of the nonstationary Schrodinger equation for the wave function of the joint system consisting of all impurity atoms and the electromagnetic field, including vacuum reservoir. Full Hamiltonian $\widehat{H}$ of the joint system can be presented as a sum of Hamiltonian $\widehat{H}_0$ of noninteracting atoms and variable field and operator $\widehat{V}$ of their interaction. The influence of the external constant electric field and static internal fields of the dielectric matrix is taken into account by shifts of the atomic energy levels.

We seek the wave function as an expansion in a set of eigenfunctions of the operator $\widehat{H}_{0}$. Using this representation of the wave function, we convert Schrodinger equation to the system of linear differential equations for the amplitudes of the quantum states. The total number of equations in this system is equal to infinity.

The key simplification of the approach is in the restriction of the total number of states taken into account. We assume that the initial excitation is weak, and all nonlinear effects are negligible. With the accuracy up to the second order
of the fine structure constant, we can consider only the states with no more than one photon (see \cite{Stephen_1964}).

Despite the restriction of the total number of quantum states, the set of equations remains infinite because of the infinity number of the single-photon field states. We can, however, formally solve it without any additional approximations. For this purpose we express the amplitudes of the quantum states with single photon via the amplitudes of the states corresponding to atomic excitation without photons. Then we put these expressions in the equations for the amplitudes of single photon states. In this way we obtain a closed finite system of equations for the amplitudes $b_{e}$ of the quantum states with one excited atom in the ensemble.

For Fourier components $b_{e}(\omega)$ we have
(at greater length see \cite{KS_JETP_2016, SKH_JETP_2011})
\begin{equation}\label{2}
\sum_{e'}\bigl[(\omega-\omega_{e})\delta_{ee'}-\Sigma_{ee'}(\omega)\bigl]b_{e'}(\omega)=i\delta_{e s}.
\end{equation}

When deriving this expression, we assumed that at the initial time only one atom is excited (this state is denoted by index $s$), while all other atoms are in the ground state.  The electromagnetic field at $t=0$ is in the vacuum state. The index $s$ as well as the indexes $e$ and $e'$ contain information both about the number $a$ of atom  and about specific atomic sublevel excited in the corresponding state.

The matrix $\Sigma_{ee'}(\omega)$ describes both spontaneous decay and photon exchange between the atoms. It plays a key role in
the microscopic theory. The explicit expressions for the elements of this matrix corresponding to a Fabry-Perot cavity were derived in \cite{KS_JETP_2016, KS_PRA_2016}.

The size of the system (\ref{2}) is determined by the number of atoms $N$ and the structure of their energy levels. In this paper
we consider the impurity atoms with the ground state $J=0$. Total angular momentum of the excited state is $J=1$. It
includes three Zeeman sublevels $|J, m\rangle$, which differ by the value of angular momentum projection on the quantization axis $z$: $m =-1,0,1$. Therefore, the total number of onefold atomic excited states is $3N$. Further in this paper we assume that the quantization axis $z$ is directed perpendicularly to the charged mirror and, consequently, along its electrostatic field.

Due to the external electrostatic field $\bf{\mathcal{E}}$ of a charged plate and internal random field of the dielectric medium, resonant frequencies of different atomic transitions $\omega_{a_m}$ differ from those of an isolated atom in free space $\omega_{0}$.
\begin{equation}\label{3}
\omega_{a_m}=\omega_{0}+\Delta_{a_m}+\Delta\omega_{m},
\end{equation}
where $\Delta_{a_m}$ is the frequency shift of the sublevel $m$ of atom $a$ ($a = 1,...,N$), which depends on its spatial position due to inhomogeneity of internal fields in dielectric; and $\Delta\omega_{m}$ is Stark shift caused by the electrostatic field of a plate, which is the same for similar transitions of different atoms.

Hereafter in this paper we will assume that $\Delta_{a_m}$ is Gaussian random variable with zero mean value and RMS deviation $\delta$, and its distribution does not depend on $m$. We denote the Stark shift of the resonant frequency of the transition $J=0$ $\leftrightarrow$ $J=1, m=\pm1$ as $\Delta\omega_{m=\pm1}$; for the transition $J=0$ $\leftrightarrow$ $J=1, m=0$ it is $\Delta\omega_{m=0}$ . The influence of an electrostatic field on the character of photon exchange is significant in the case when Stark splitting $\Delta=\Delta\omega_{m=0}-\Delta\omega_{m=\pm1}$ is comparable with the natural linewidth $\gamma_{0}$ of an isolated atom.

Numerical solution of the system (\ref{2}) allows us to obtain the Fourier amplitudes of atomic states $b_{e}(\omega)$. Using $b_{e}(\omega)$ we
can obtain the amplitudes of all states taken into account in our calculations (see \cite{KS_JETP_2016, SKH_JETP_2011}) and, consequently, the wave function of the considered system.

To analyze the dynamics of atomic ensemble located near a single mirror on the basis of mathematical formalism developed for a cavity, we should go to the limit of infinite distance between the mirrors and consider atoms near the first mirror. In this case the influence of the second mirror on the dynamics of atomic system can be neglected.

Note that any physical observables that we will analyze depend on the positions of all impurity atoms. In this paper we consider spatially disordered atomic ensembles with uniform (on average) distribution of atomic density, as it is the case in experiments. By this reason we average all the results over random spatial configurations of the ensemble by a Monte Carlo method. To take into account the inhomogeneous broadening we also perform Monte Carlo averaging over random shifts $\Delta_{a_m}$ of energy levels caused by the inhomogeneity of the internal fields of a dielectric.

In the next section, we use the general approach to investigate the simultaneous influence of the peculiarities of the spatial structure of field modes near the conductive surface as well as Stark splitting of energy levels induced by an electrostatic field on the character of many-body cooperative effects. We will calculate the transition spectrum of an excited atom surrounded by an ensemble of unexcited atoms, and spontaneous decay dynamics. On this basis, we will analyze the influence of the electrostatic field on radiation trapping in the considered system.

\section{Results and discussion}
Part of the effects, caused by the influence of a charged conducting surface on an atomic ensemble, can be described within the framework of monatomic approximation and appear for dilute ensembles or even for single atom. Some effects, caused by the modification of the interatomic dipole-dipole interaction due to simultaneous influence of the conducting surface and the electrostatic field, are essentially collective.

Monatomic effects are relatively simple and have been well studied to date. When a single atom is located close to the uncharged surface, the spectrum of atomic transition represents a Lorentz profile, like in the case of a free atom. But the linewidth $\gamma$ differs from one of a free atom and depends on the distance $z$ between the atom and the surface. If $z$ is less or comparable with the resonant wavelength $\lambda_0$, the difference is very significant. Accordingly, the dynamics of the spontaneous decay of the excited atom is described by a single-exponential law, $P_{s}(t)=\exp(-\gamma t)$. The function $\gamma(z)$ depends on Zeeman sublevel, which is initially populated. Thus, for Zeeman sublevels $m=\pm 1$,  $\gamma(z)$ converges to zero if atom approaches to surface. For $m=0$ this limit is equal to $2\gamma_0$.  As $z$ increases both values tend to $\gamma_0$. On the whole, the function $\gamma(z)$ has a nonmonotonic oscillating character (see for example \cite{KS_JETP_2018}).

If the conducting surface is additionally charged, its electric field causes Stark shifts which actually does not influence the monatomic effects. Only the frequencies of atomic resonances change. Their shapes remain the same. The amplitude and the width of the resonance change absolutely negligibly because the Stark shift is absolutely negligible in comparison with the frequency of any optical transition.

Collective effects in dense atomic ensembles under considered conditions have been studied in less detail. We begin our analysis with studying  the shape of atomic transition connected with spontaneous decay of an atom initially excited in dense atomic ensemble. We assume that at the initial time all the other atoms of the ensemble are unexcited.

\subsection{Atomic transition spectrum}

As it is clear from the aforesaid, the effect of the surface depends on the positions of all atoms and especially of the excited one $z_{exc}$. The most interesting phenomena are observed if $z_{exc}$ does not exceed the wavelength of resonant light. By this reason, further we will consider
$z_{exc}=1$  assuming that reference point $z=0$ corresponds to the position of the surface (hereafter, we take $\lambdabar=k_{0}^{-1}=\lambda_{0}/2\pi$ as the unit of length). Also for simplicity we assume at first that inhomogeneous broadening is negligible, so that $\Delta_{a_m}=0$ (respectively, $\delta=0$). In this case all the atoms are resonant to each other, so the role of the dipole-dipole interaction is manifested to the maximum extent.

In the general case, the specific type of transition spectrum for a given density depends not only on $z_{exc}$ but also on the size of the atomic ensemble. We have previously analyzed size dependence of the transition spectrum. When the size is comparable with the mean free path of a photon, the changes of the transition spectrum with increasing in size are essential. As linear size increase, these changes become more and more weak. Size dependence has an evident tendency to saturation. Further we present the results, which correspond to sufficiently large sample, when size dependence can be neglected. So it can be used for a description of the transition spectrum of excited atom inside any macroscopic ensemble with reasonable accuracy.

\begin{figure}\center
	\includegraphics[width=7cm]{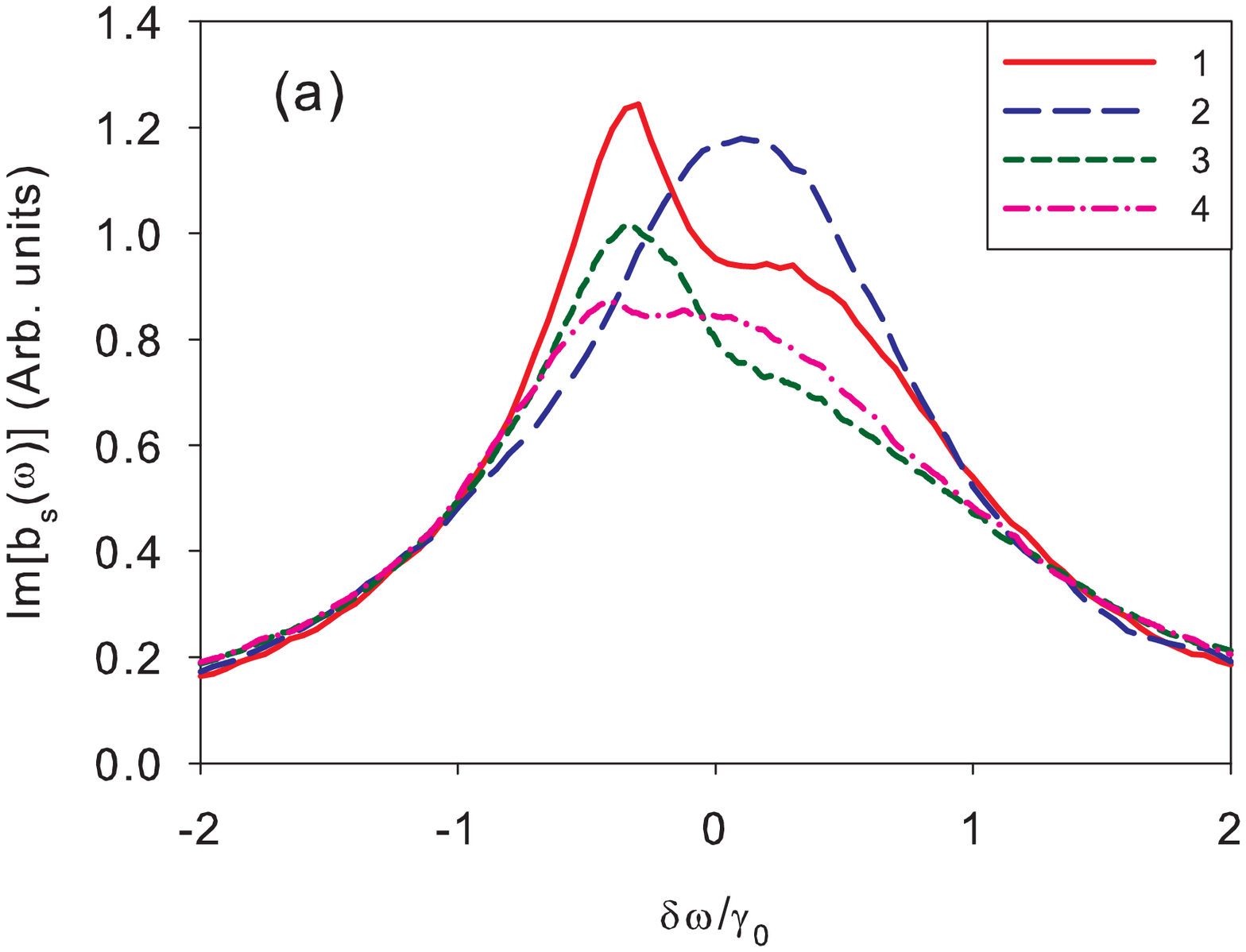}
	\includegraphics[width=7cm]{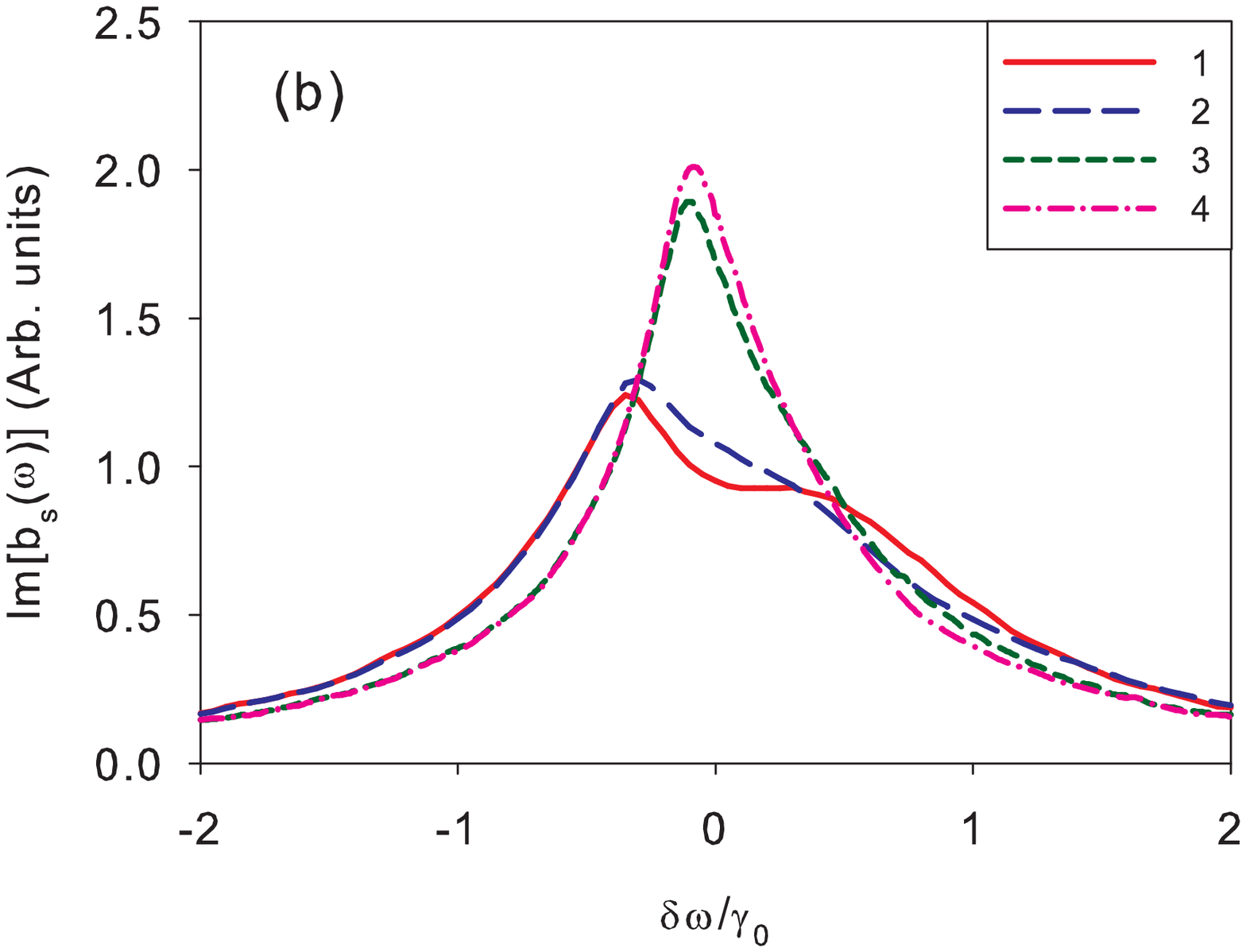}\\
	\caption{\label{fig:one}
			Transition spectrum of an atom inside  atomic ensemble with $n=0.05$, $\delta=0$. (a) $m=0$; (b) $m=\pm1$. 1, Electric field is absent $\Delta=0$ and there is no surface; 2, Electric field resulting $\Delta=\gamma_{0}$; 3, There is a conducting surface  $z_{exc}=1$ and electric field is absent; 4, There are both the field resulting $\Delta=\gamma_{0}$ and the surface. The detuning $\delta\omega$ is calculated from the resonant frequency taking into account Stark shift (if it is nonzero.)}\label{f1}
\end{figure}

The line shape of atomic transition corresponding to the decay of Zeeman sublevel $m=0$ is shown in the Figure \ref{fig:one}(a).  Here we compare the shape of atomic resonance in four cases. For convenience of the comparison, the frequency  is calculated from the resonant frequency taking into account Stark shift $\delta\omega=\omega-\omega_{a_m}$ (see Eq. (\ref{3})). The first curve is obtained  when both the surface and electric field are absent. The specific dimensionless atomic density is chosen equal to $n=0.05$. We see that it is sufficiently large value, so that the dipole-dipole interaction plays an important role and the shape is essentially different from Lorenz contour typical for a free atom. Curve 1 transforms into curve 2 when we switch on the electric field corresponding to the Stark splitting $\Delta=\gamma_{0}$. Electric field without conducting surface causes essential shift and essentially modify the shape of the resonance. Here we see the influence of electric field on collective effects, partially on collective Lamb shift caused by modification of resonant dipole-dipole interatomic interaction. Placing the atomic ensemble near uncharged surface (curve 3 in Fig. \ref{fig:one}(a)) change amplitude but practically does not transform the shape of the resonance and the collective shift.  Simultaneous  influence of the electric field and surface causes the change of the collective Lamb shift, width of the resonance as well as its shape.

The presence of electric field and/or the surface makes the system optically anisotropic. By this reason their influence on the shape of the transitions $J=0$ $\leftrightarrow$ $J=1, m=\pm1$ differs from that corresponding to the transition $J=0$ $\leftrightarrow$ $J=1, m=0$. It can be seen in Fig. \ref{fig:one}(b).  The surface causes essential narrowing of the resonance which is explained mainly by monatomic effects. For $z_{exc}=1$, $\gamma_{m=\pm 1}=0.65\gamma_0$. The effect of the electric field is weak for considered transition, which agrees with the previously obtained results of the calculation of the dielectric constant tensor \cite{sokolov17, sokolov18}. Note however that near the surface different Zeeman sublevels not only decay in different ways, but also perceive the effects of an electric field in different ways. This can be understood if we compare Fig. \ref{fig:one}(b) with Fig. \ref{fig:one}(a).

\begin{figure}\center
	\includegraphics[width=7cm]{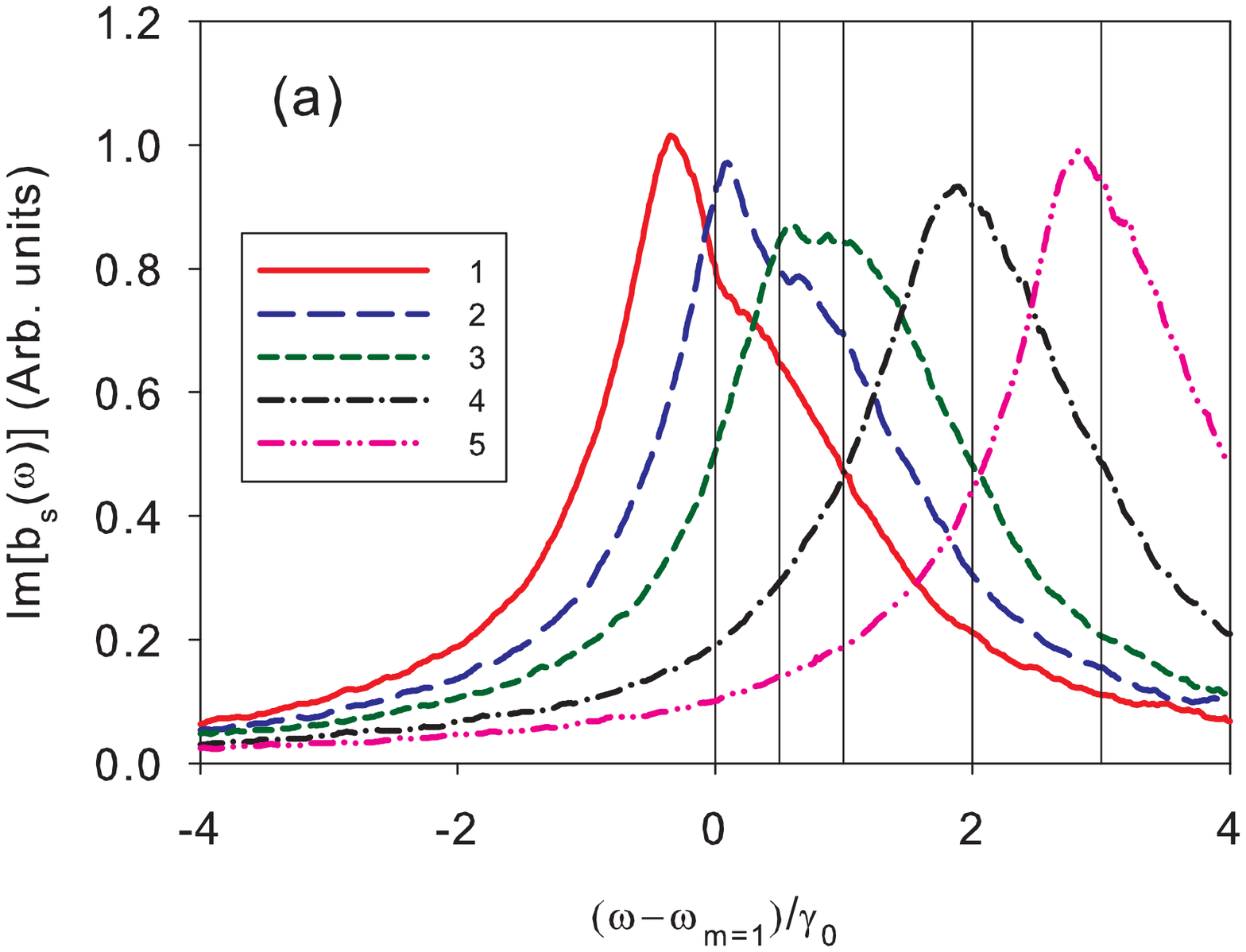}
	\includegraphics[width=7cm]{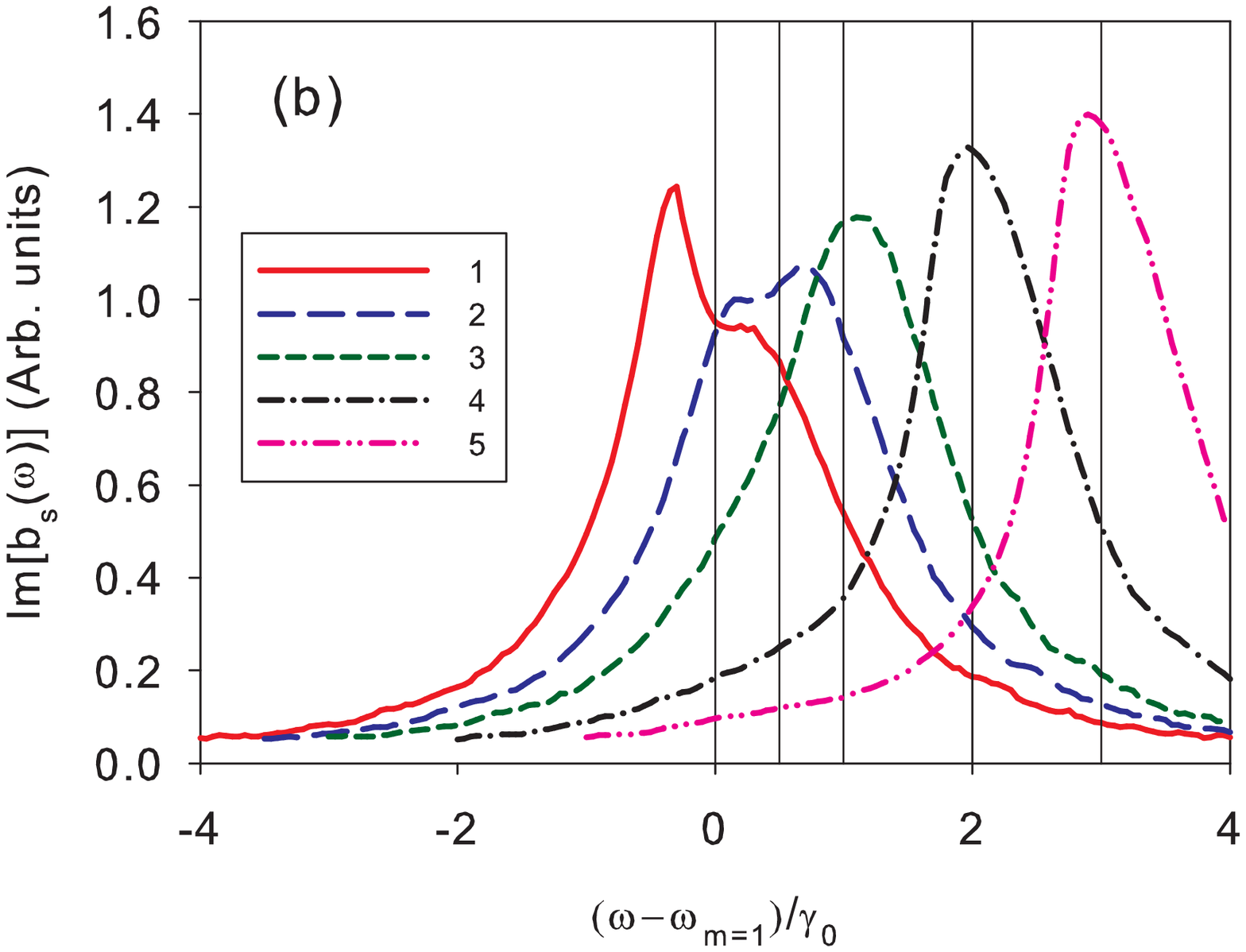}\\
	\caption{\label{fig:two}
			Transition spectrum of an atom in electric field. $m=0$, $n=0.05$, $\delta=0$. (a) Near a conducting surface, $z_{exc}=1$; (b) in free space. 1, $\Delta=0$; 2, $\Delta=0.5\gamma_{0}$; 3, $\Delta=\gamma_{0}$; 4, $\Delta=2\gamma_{0}$; 5, $\Delta=3\gamma_{0}$.}\label{f2}
\end{figure}

The influence of the electric field changes with its magnitude. This dependence is most pronounced for transition $J=0$ $\leftrightarrow$ $J=1, m=0$. It is illustrated by the Figure \ref{fig:two} where we show the shape of the transition spectrum for different Stark splitting $\Delta$. For clarity, in Fig. \ref{fig:two} the frequency is calculated from the resonant frequency of the transition $J=0$ $\leftrightarrow$ $J=1, m=\pm1$ of a free atom taking into account the Stark shift, $\omega_{m=1}=\omega_{0}+\Delta\omega_{m=\pm1}$. Fig. \ref{fig:two}(a) and Fig. \ref{fig:two}(b) correspond to the cases with and without conducting surface. In these figures we added reference vertical lines, which indicate all the considered values of Stark splitting.

In the case of the atomic ensemble near the surface, the parameters of the resonance such as its shape, amplitude and width change with magnitude of constant field non-monotonously. The collective Lamb shift is also nonmonotonic. The maximal distortion corresponds to the splitting which is close to the natural linewidth of a free atom.

The influence of the electric field on collective effects in free space (see Fig. \ref{fig:two}(b)) has some peculiarities. We see very strong distortion of the resonance shape for Stark splitting less than $\gamma_0$. For strong field, corresponding to $\Delta>\gamma_0$, the collective Lamb shift is less than one in the case of atomic ensemble near the surface (compare with Fig. \ref{fig:two}(a)). Besides that, increasing of the field causes some line narrowing and increasing of the amplitude of the resonance.

In the Fig. \ref{fig:two} we show the line shape up to Stark splitting equal to $\Delta=3\gamma_0$. It is clear that the dependence of the observable spectrum on the Stark splitting $\Delta$ should disappear when this splitting becomes more than atomic level shifts caused by resonant dipole-dipole interaction. Our calculation indicate that for a considered density it takes place at $\Delta \sim 15\gamma_0$.

In conclusion of this section, we note that the solution of an algebraic system of equations (\ref{2}) with a given right-hand side is equivalent, in essence, to finding its Green function with given point source. Calculation of the amplitude $b_s(\omega)$ means determination of the Fourier component of the Green function in the point of the source. In accordance with \cite{prl09}, knowledge of this function allows us to determine the local density of states of atomic system, as well as to find a number of characteristics of this system, for example, the mean free path of photons inside it.

\subsection{Light trapping}

The influence of the charged conducting surface on an atomic ensemble can be detected in the experiment, for example, by measurement of its afterglow after initial excitation. The dynamics of the total intensity and, consequently, light trapping is determined by the dynamics of the atomic excited state population.

In this subsection we analyze time dependence of the total population of the excited states of all atoms of the ensemble. As earlier, for simplicity we assume that initially only one atom is excited. Our analysis is based on the calculation of the inverse Fourier transform of $b_{e}(\omega)$. It allows us to obtain
the time dependence of the quantum amplitudes of the onefold atomic excited states, $b_{e}(t)$. The time-dependent
population of any Zeeman sublevel of any atom in an ensemble can be calculated in a standard way: $P_{e}(t)=|b_{e}(t)|^{2}$.

\begin{figure}\center
	\includegraphics[width=7cm]{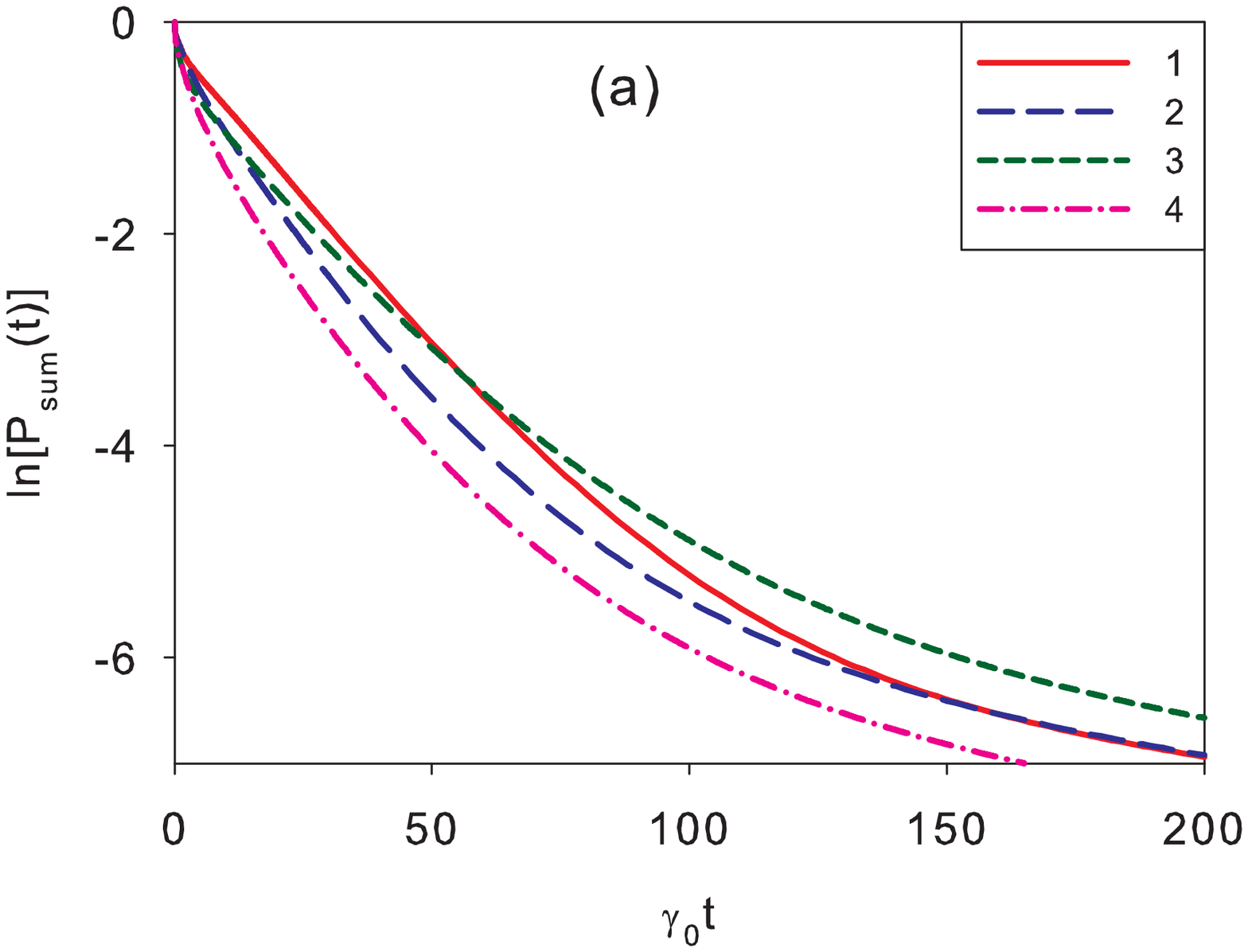}
	\includegraphics[width=7cm]{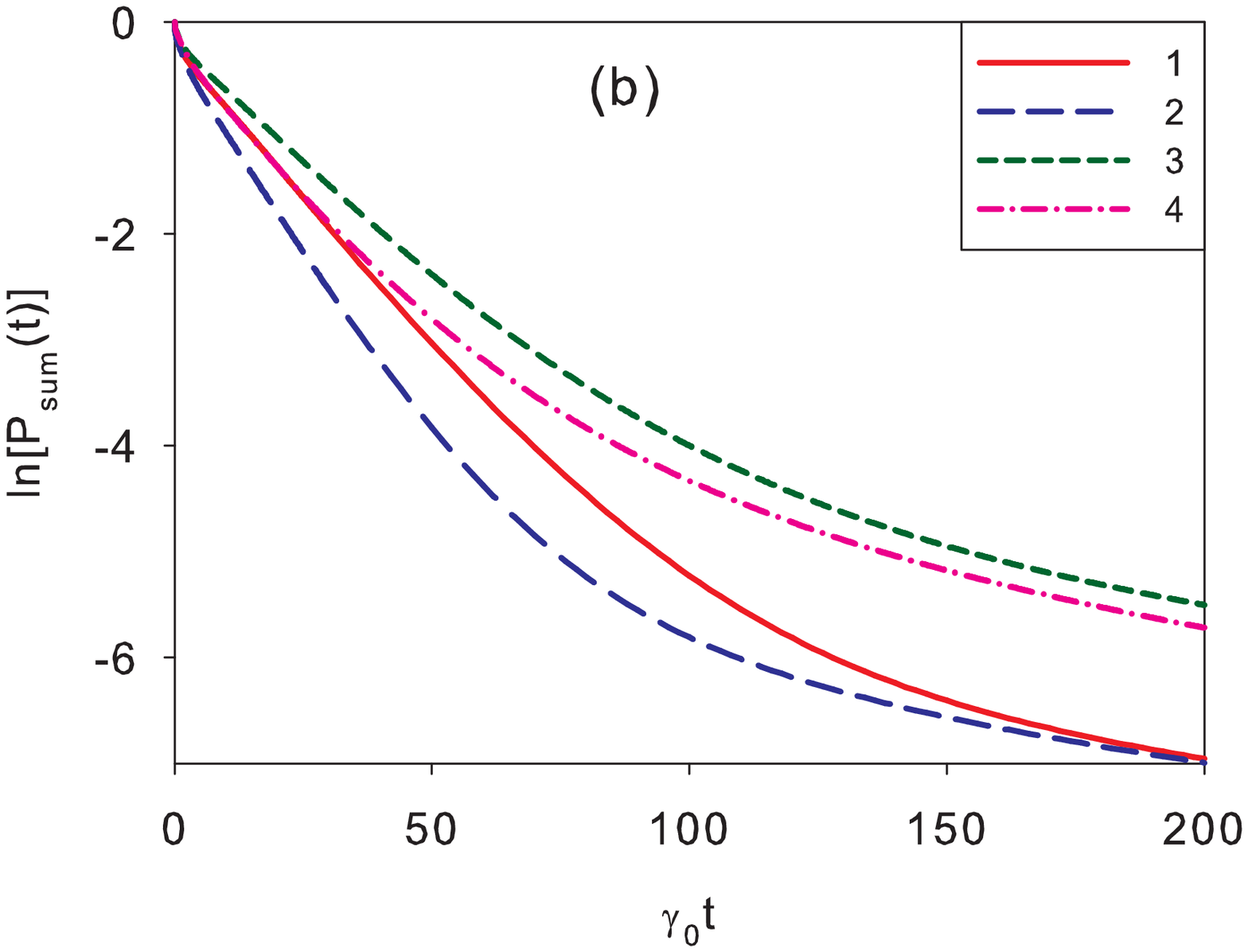}\\
	\caption{\label{fig:three}
			Time dependence of the total excited state population of atomic ensemble with $n=0.05$, $\delta=0$. (a) $m=0$; (b) $m=\pm1$. 1, Atomic ensemble in free space 2; Electric field resulting $\Delta=\gamma_{0}$, there is no surface; 3, There is a conducting surface  $z_{exc}=1$ and electric field is absent; 4, There are both the field resulting $\Delta=\gamma_{0}$ and the surface.}\label{f3}
\end{figure}

The total excited state population $P_{sum}(t)$ is given by a sum of $|b_{e}(t)|^{2}$ over all atoms in the ensemble. Figure \ref{fig:three} shows the time dependence of the total excited state population in the case $\delta=0$. The results are presented for the atomic ensemble of cylindrical shape, one of the planes of a cylinder coincides with the plane of conducting surface, the radius of a cylinder is $R=12$, the length is $L=13$, which is much more than the mean free path of a photon at the considered density $n=0.05$. The initially excited atom is located at $z_{exc}=1$ on the central axis of a cylinder.

In Fig. \ref{fig:three}, like in Fig. \ref{fig:one}, we compare four main cases: ensemble in free space, in electric field, near the uncharged surface and near the charged surface. In all the cases we see typical manifestation of collective effects. The dynamics of an atomic excitation can not be described by a simple one-exponential law like in the case of a single atom. It is explained by interatomic interaction caused by the photon exchange between different atoms.  Among different collective quantum states formed as a result of this interaction in the considered ensemble there are both super- and subradiant ones. In such a case the spontaneous decay dynamics is described by a multi-exponential law.

Besides these typical collective effects, Fig. \ref{fig:three} demonstrate some features determined by surface and electric field. We see that both these factors separately influence the nature of the decay dynamics which is connected with mentioned above modification of the dipole-dipole interatomic interaction and, consequently, with changes in both sub- and superradiant states.

It should be noted that these factors, when combined, can strengthen each other or compensate. Thus, for $m = 0$, their combined effect accelerates the decay of excitation. Curve 4 in Fig. \ref{fig:three}(a), which describes the dynamics of the decay of the excitation near the charged surface, decreases most rapidly. In the case of the initial excitation of an atom to the level $m = \pm 1$, on the contrary, both factors act in different directions, and curve 4 in Fig. \ref{fig:three}(b) demonstrate intermediate decay rate.

Comparison of the curves 1 and 2, as well as 3 and 4, makes it possible to reveal the influence of the electric field on the nature of the afterglow of the ensemble in the absence of a conducting surface and near it. This comparison gives grounds to conclude that a change in the structure of the modes of the electromagnetic field, caused by the presence of a surface, leads to a change in the effect of the constant field, i.e. modifies electro-optical effects in dense atomic systems.

\begin{figure}\center
	\includegraphics[width=7cm]{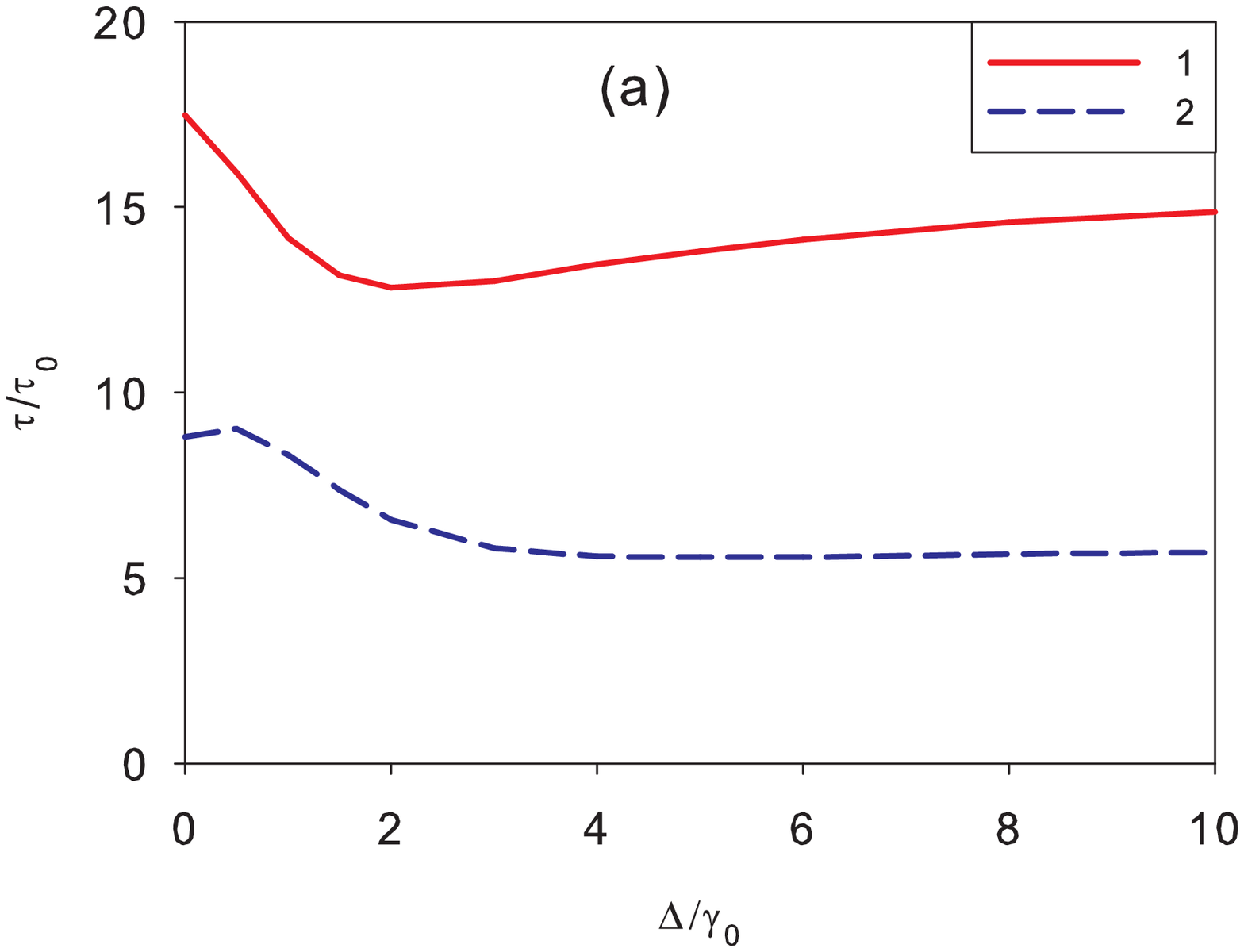}
	\includegraphics[width=7cm]{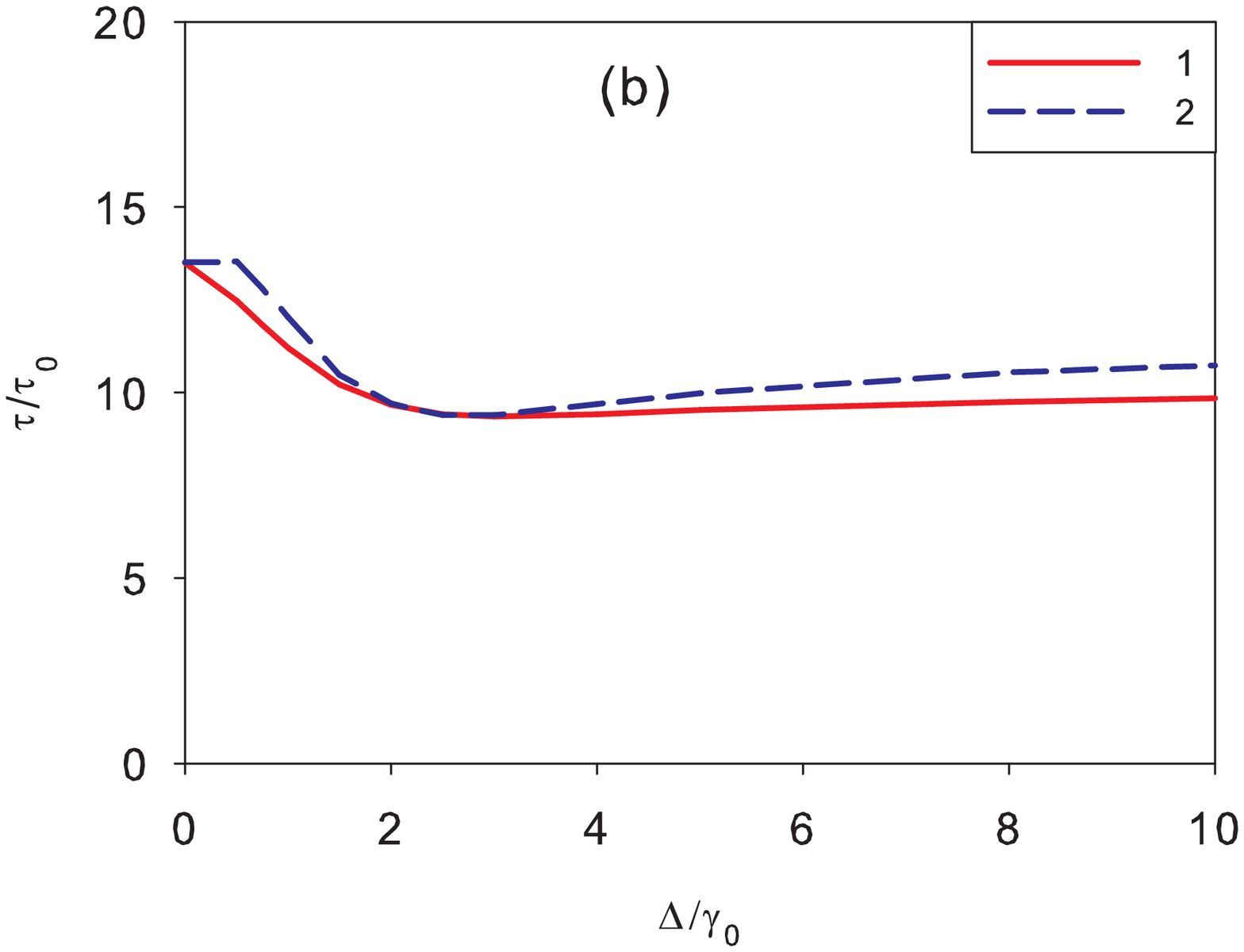}\\
	\caption{\label{fig:four}
	The time of radiation trapping depending on the electrostatic field strength. $n=0.05$, $\delta=0$. (a) Atomic ensemble near the conducting surface; (b) Atomic ensemble in free space. 1, $m=\pm1$; 2, $m=0$. $\tau_{0}=1/\gamma_{0}$ is the natural lifetime of the excited states of a free atom.}\label{f4}
\end{figure}

This circumstance is also confirmed by the analysis of the typical time of radiation trapping. We will estimate this time $\tau$ from the relation $P_{sum}(\tau)=1/e$. Figure \ref{fig:four} demonstrates, how this time changes with increasing in the electrostatic field strength both for an ensemble in free space (Fig. \ref{fig:four}(a)) and near the conducting surface (Fig. \ref{fig:four}(b)). The field strength, as earlier, we characterize by the Stark splitting $\Delta$.

In the Fig. \ref{fig:four} we see that the electric field can significantly affect the light trapping. The detailed analysis shows, that the dependence of the time of radiation trapping on the Stark splitting is complex, in some cases it can even be nonmonotonic. So in the Fig. \ref{fig:four} we see that this dependence predominantly decreases in the diapason from $\Delta=0$ up to $\Delta\sim3\gamma_{0}$. With further increasing of $\Delta$, the time of radiation trapping slowly increases.

\begin{figure}\center
	\includegraphics[width=7cm]{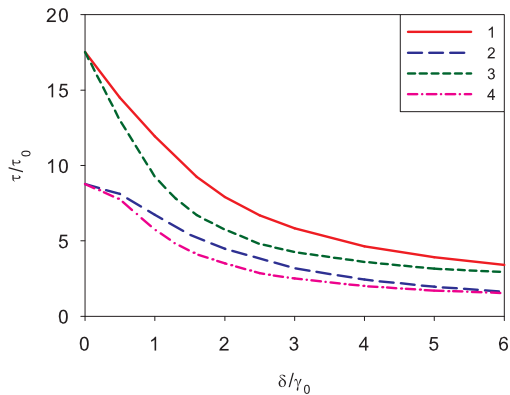}\\
	\caption{\label{fig:five}
			The time of radiation trapping depending on the inhomogeneous broadening. 1, 2, $n=0.1$; 3, 4, $n=0.05$. 1, 3, $m=\pm1$; 2, 4, $m=0$.}\label{f5}
\end{figure}

To conclude this section of the article, consider the influence the inhomogeneous broadening caused by internal fields of the dielectric $\delta$ on the light trapping. Dependence of the trapping time $\tau$ on $\delta$ for different initial conditions of excitation is shown in Fig. \ref{fig:five}.
The calculations were performed for $\Delta=0$. As $\delta$ increases, the mean free path of the photon also
increases, because the role of cooperative multiple scattering becomes weaker. This leads to a monotonic decrease of the time of radiation
trapping.

Note, however, that the suppression of collective effects with increasing in $\delta$ manifests itself slower as the density of impurity centers increases. Increasing of the density compensates negative influence of the inhomogeneous broadening on the collective effects. This fact is  confirmed by comparison of the curves shown in Fig. \ref{fig:five} which correspond to different atomic densities. Therefore, for high densities of impurities, very often used in experiments, influence of the dipole-dipole interaction can be significant even for large inhomogeneous broadening, $\delta\gg\gamma_0$.

When the sizes of a sample are fixed, increasing of the density obviously leads to an increase in the trapping time. For us now, not the absolute value of the trapping time is important, but its decay rate with increasing in $\delta$. For this reason, considering the ensembles of the density $n=0.1$, we reduced the size so that for $\delta=0$ the absolute value of the trapping time is the same as one for the density $n=0.05$. Comparison of the curves 1 and 3, as well as 2 and 4 shows that with density increasing, the mutual non-resonance of different impurities centers becomes less pronounced.

\section{Conclusion}
We have studied many-body cooperative effects caused by the dipole-dipole interaction in an ensemble of pointlike impurity centers imbedded into transparent dielectric and located near a charged perfectly conducting surface.  On the basis of the
general quantum microscopic theory, we have analyzed the simultaneous influence of the surface and the electrostatic field on the transition spectrum of an excited atom inside an ensemble, as well as on the dynamics of the total excited state population related to the whole ensemble. The cooperative Lamb shift depending on the electric field has been studied. The time of radiation trapping as a function of the electric field strength and the inhomogeneous broadening has been investigated.

In our opinion, of special interest is the application of the theory described in the present paper for the investigation of Anderson localization
of light in
quasi-two-dimensional ensembles of impurity centers
embedded into a transparent dielectric and located
near a charged conducting surface. This is associated
with the fact that, in the systems of reduced dimensionality,
cooperative phenomena have a number of nontrivial
features that promote the Anderson
localization. Moreover, the electrostatic field partially removes
the degeneracy of the multiplet of the excited state,
which additionally contributes to
the strong localization of light \cite{Skipetrov_2015}. Despite the absence of the Anderson localization in atomic ensembles in free space, even in the presence of the electric field, as it has been proved in \cite{Skipetrov_2019}, the combined effect of the surface and the electric field gives us hope to detect the Anderson localization.

One more promising direction for the development of the theory described in the present paper is its generalization to the analysis of the dipole-dipole interaction in atomic ensembles placed in a waveguide.
The case when the resonant frequency of atomic transition
is less than the cutoff frequency of the waveguide attracts
particular interest due to spontaneous decay suppression of all
the Zeeman sublevels. Moreover, the analysis of the atomic
systems in a waveguide can be useful for the investigation of
Anderson localization, because in quasi-1D systems all the
collective states are localized \cite{Akkermans_2013}, \cite{LesHoushes}.

\section*{Acknowledgments}

This work was partially supported by Russian Foundation for Basic Research (Grant No. 18-32-20022) and the Foundation for the
Advancement of Theoretical Physics and Mathematics ”BASIS”. The calculation of light trapping (part III B)
was supported by the Russian Science Foundation (Grant No. 17-12-01085).

\end{document}